# Designing electromagnetic resonators with quasinormal modes

**Tong Wu[1][†], Philippe Lalanne[1]**

[1]Laboratoire Photonique, Numérique et Nanosciences (LP2N), IOGS- Université de Bordeaux-CNRS, Talence, France

[†]wutong1121@gmail.com

## Abstract

Micro- and nanoresonators, which enable light trapping in small volumes for extended durations, play a crucial role in modern photonics. The optical response of these resonators is determined by their fundamental resonances, known as quasinormal modes (QNMs). Over the past decade, the electromagnetic theory of QNMs has undergone significant development and has now reached a level of maturity that allows its reliable application to numerous contemporary electromagnetic problems. In this review, we explore recent applications of QNM theory for designing and understanding micro and nanoresonators. We highlight why QNMs provide deep physical insights and enhance computational efficiency in scenarios involving mode hybridization and perturbation.

## 1 Introduction

Micro or nanoresonators play a crucial role in advancing modern photonics [1,2]. The interaction of light with optical resonators is fundamentally governed by the excitation of intrinsic natural resonant modes. When excited by a pulse, these modes initially store energy and subsequently release it through exponential decay. Known as quasinormal modes (QNMs), they are source-free solutions, $[\widetilde{\mathbf{E}}_m(\mathbf{r}), \widetilde{\mathbf{H}}_m(\mathbf{r})]\exp(-i\widetilde{\omega}_m t)$ to Maxwell's equations with complex frequencies $\widetilde{\omega}_m$

$$\nabla \times \widetilde{\mathbf{E}}_m = i\widetilde{\omega}_m \mu_0 \widetilde{\mathbf{H}}_m, \nabla \times \widetilde{\mathbf{H}}_m = -i\widetilde{\omega}_m \boldsymbol{\varepsilon}(\mathbf{r}, \widetilde{\omega}_m) \widetilde{\mathbf{E}}_m, \qquad (1)$$

and satisfy the outgoing-wave condition for $|\mathbf{r}| \to \infty$. Hereafter, $\widetilde{\mathbf{E}}_m$ and $\widetilde{\mathbf{H}}_m$ respectively denote the *normalized* electric and magnetic fields [3,4], $\boldsymbol{\varepsilon}$ denotes the possibly dispersive permittivity tensors. Unlike the normal modes of Hermitian systems, which have real frequencies, the imaginary part $\text{Im}(\widetilde{\omega}_m)$ is non-zero, accounting for absorption or radiation losses [3].

The QNMs of electromagnetic resonators are characterized by two main quantities: their mode volume ($\widetilde{V}_m$) and quality factor ($Q_m$) [4,5]. The former is related to the spatial extent of the electromagnetic field and the latter is proportional to the confinement time in units of the optical period, $Q_m = -(1/2)\text{Re}(\widetilde{\omega}_m)/\text{Im}(\widetilde{\omega}_m)$ [3,5,6].

High-$Q$ resonators trap light for long time, whereas small mode volume resonators confine light in tiny volumes. Resonators with large $Q_m/\widetilde{V}_m$ ratios strongly enhance the interaction between the trapped photons and the host materials, giving rise to significant nonlinear, quantum and optomechanical effects [1-4].

Photonic applications generally rely on two types of structures which respectively offer high $Q_m$ and small $\tilde{V}_m$. High-$Q$ resonators, typically with $Q_m \sim 10^6$, are usually fabricated using lossless dielectric materials, such as the photonic crystal microcavities (Figure 1a). These structures can confine light for extended periods of time (typically $Q_m$ cycles) within volumes of about the resonant wavelength cube $\lambda_R^3$.

Resonators with mode volumes significantly smaller than $\lambda_R^3$ are engineered using metallic nanostructures that support localized plasmons [7]. One notable example is the nanoparticle-on-mirror (NPoM) construct (Figure 1b) [2,8], which represents a current area of research interest. NPoM trap light in nanometer-sized dielectric gaps between two metal surfaces, achieving mode volumes as small as $\sim 10^{-7}\,\lambda_R^3$. However, this exceptional spatial field confinement comes with a drawback: the inherent absorption of the materials restricts the quality factor to small values, $Q \sim 10 - 100$.

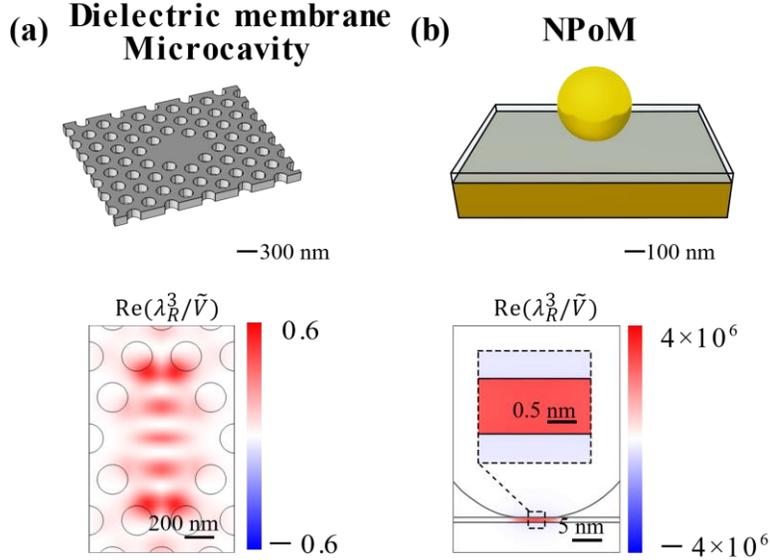

**Figure 1.** Examples of micro and nanocavities. (a) A photonic-crystal microcavity with a 330nm lattice constant in a 320nm-thick dielectric membrane has typical resonance wavelengths $\lambda_R$ at 1.5 µm. (b) A metal nanoparticle on mirror (NPoM) structure consists of a faceted gold nanoparticle (radius R=20 nm) separated from a gold substrate by a 1nm-thick dielectric film. The maps depict the real part of the inverse mode volume, $\text{Re}(1/\tilde{V}_m)$. The microcavity confines light within approximately $\lambda_R^3$, whereas the NPoM achieves a minimum mode volume of about $10^{-7}\,\lambda_R^3$. Note that the imaginary part, $\text{Im}(1/\tilde{V}_m)$, of the mode volume is not shown. Adapted from [5].

QNMs can be used to expand the electromagnetic field $\mathbf{E}^s(\mathbf{r},\omega)\exp(-i\omega t)$ scattered by resonators illuminated by monochromatic waves at frequency $\omega$ [3,9]

$$\mathbf{E}^s(\mathbf{r},\omega) = \sum_m \alpha_m(\omega)\tilde{\mathbf{E}}_m(\mathbf{r}), \qquad (2)$$

where the $\alpha_m$'s are modal excitation coefficients. The latter are known analytically and can be calculated as a spatial overlap between the normalized QNM field and the incident wave. They describe the contribution of the resonance modes to the optical response. A time-domain QNM expansion formula can also be derived from Equation (2) by performing a Fourier transformation from the frequency domain to the temporal domain

$$\mathbf{E}^s(\mathbf{r},t) = \text{Re}[\sum_m \beta_m(t)\tilde{\mathbf{E}}_m(\mathbf{r})]. \tag{3}$$

In the past decade, there have been significant advancements in electromagnetic QNM theory. These include resolving the critical issue of QNM normalization [10-15], testing completeness of the QNM expansion across various systems [10,16-23], and extending the analysis to the temporal domain [20,24-27].

Nowadays, QNM theory is extensively utilized in designing optical resonators for various applications [28-30], including second- and higher-harmonic generation [31-33], optical parametric oscillators [34], Bell state generation [35], random lasing [36,37], cavity QED [38-40], chiral molecule sensing [41,42], quantum plasmonics [43-45], structural color generation [46], visual appearance generation [47], random medium [48], and ultrafast optics [24,27]. But why are QNMs essential for nanoresonator design?

One explanation lies in the unique physical insights provided by QNM theory, which are often unattainable from brute-force numerical simulations. While the latter can be used to accurately compute the resonant spectra of resonator responses [2,49-52], the spectrum interpretation can be indirect and sometimes incomplete. For instance, far-field incident light might not efficiently excite all modes, causing dark modes to be overshadowed by bright ones in the spectra. Additionally, spectrally overlapping resonances are difficult to distinguish, even with semi-analytical tools like the temporal-coupled mode theory [53,54]. These resonances might merge to form a complex Fano response that appears as a single bell-like response, potentially leading to an incorrect interpretation with a single resonance [11,55].

In contrast, QNMs are intrinsic to the system and independent of the incident field. By computing QNMs, optical dark modes can be identified unambiguously [56-60], and spectrally overlapping modes can be distinguished [11,55,61]. QNM expansion methods allow the reconstruction of optical scattering spectra [3,9,14,20,62,63] with a weighted sum of QNMs. They may also provide explicit formula for the local density of electromagnetic states [11,62,64-67], a quantity of prime interest to interpret the optical response of resonators coupled with quantum emitters.

Another explanation is that electromagnetic QNM theory may also offer a significant improvement in computational efficiency over classical modeling tools that operate in the real-frequency or temporal domains. In frequency-domain simulations, calculations are repeated for each frequency, while in time-domain simulations, they must be repeated for different excitation fields.

The QNM expansion formula in Equation (2), which includes analytically-known $\alpha_m$ coefficients, allows efficient computation of the optical responses to arbitrary incident waves, once the dominant QNMs are determined [3,68]. This efficiency is particularly useful for predicting the responses of resonant structures to various incident fields [69]. Applications include calculating the bidirectional reflectance distribution function (BRDF) of disordered metasurfaces [47], and assessing optical forces [70,71], for instance.

Recent advancements in QNM perturbation [72,73] and coupled QNM-theory [74-76] enable the prediction of the QNMs of altered geometries based on the QNMs of the initial geometry. They enhance computational efficiency not only for frequency sweeps but also for varying parameters such as shape or permittivity. This is particularly valuable for inverse design problems that require optical responses over plenty of parameter spaces.

This review aims to highlight recent advancements in applying QNMs to nanoresonator design, emphasizing the benefits of using QNM theory. Special focus is placed on how QNMs provide deep physical insights and enhance computational efficiency in mode hybridization and perturbation scenarios. For readers interested in the detailed physics and mathematical properties of electromagnetic QNMs, references such as [3,5,9,10] are recommended. This article is structured as follows: a brief overview of key concepts on electromagnetic QNMs, followed by a discussion of their applications in resonator design.

## 2 QNM theory in a nutshell

### 2.1 Computation of electromagnetic QNMs

The computation of electromagnetic QNMs is now routinely performed using mode solvers for Maxwell's equations [77]. The most common method involves calculating the poles of the resonator response (either a scattering-matrix element or a component of the scattered field at a specific spatial position) by driving the system with a source emitting at complex frequencies until the response diverges. Alternatively, one can directly solve the eigenvalue problem defined by Equation (1).

A few open-source software packages dedicated to QNM computation are available [77-79], including our comprehensive released freeware, MAN [68], which implements the pole-searching method or directly solves the eigenvalue problem.

### 2.2 The QNM divergence

QNM fields exponentially grow in space far away from the resonators, typically taking the form of a leaky spherical wave, $r^{-1} \exp[i\widetilde{\omega}_m(-t + r/c)]$ as $r \to \infty$ in 3D open spaces. The spatial divergence has significantly slowed down the development of the electromagnetic QNM theory, just like in related areas, e.g. gravitational waves [80-84].

First, the spatial divergence raises difficulties in normalizing the QNMs fields [10]. Second, the spatial divergence also raises the question of whether the expansions of Equations (2) and (3) are complete or not and when completeness is achieved, for what subspace [3,85]. Indeed, note that the scattered field at real frequency always vanishes as $r^{-1}$ for $r \to \infty$; it seems unlikely that expansions relying on fields that all divergence for $r \to \infty$ may capture the special decay. In addition to the issue of incompleteness, the divergence appears to contradict our physical intuition, raising doubts about whether it 'truly corresponds to any physical reality,' as was noted long ago [86].

These issues have been addressed through extensive efforts notably over the past three decades.

Various frameworks for QNM normalization have been developed [10-15,87,88], and the completeness of QNM expansions (possibly augmented by numerical modes arising from QNM regularization [16,20]) has been verified analytically and numerically in numerous examples [10,16-23]. A recent review [10] provides advanced details and traces historical errors.

One widely adopted normalization framework is the PML-regularization method [11]. In this approach, the continuous Maxwell operator from Equation (1) is replaced by a linear operator within a finite physical domain bounded by perfectly matched layers (PMLs). The latter maps infinite open spaces into regularized Hilbert spaces, by converting the exponential growth of QNM fields in open space to an exponential decay within the PMLs. The regularized QNMs become square-integrable and are normalized with a volume integral over the physical domain Ω inside the PML and the PML domain $\Omega_{PML}$

$$\iiint_{\Omega \cup \Omega_{PML}} \widetilde{\mathbf{E}}_m \cdot \frac{\partial \omega \varepsilon}{\partial \omega} \widetilde{\mathbf{E}}_m - \widetilde{\mathbf{H}}_m \cdot \frac{\partial \omega \mu}{\partial \omega} \widetilde{\mathbf{H}}_m d^3 \mathbf{r} = 1. \tag{4}$$

With this normalization, analytical expressions for the modal expansion coefficients can be derived from first-principle calculations [11,20]. Moreover, the QNM expansion augmented by numerical modes is complete for all **r** within the regularized space, including the PML domain. For non-dispersive resonators, biorthogonality warranties the uniqueness of the expansion coefficients $\alpha_m$. In the presence of dispersion, formulas of $\alpha_m$ depend on the choosing of the auxiliary field used for linearization and different methods for splitting the source term in Maxwell equation. Once these factors are defined $\alpha_m$ is also uniquely determined [10,18,20].

The PML regularization offers a framework to mimic the Hermitian system, by ensuring completeness and eliminating divergence. However, it does not offer a clear physical interpretation of the implications of the impact of QNM spatial divergence when analyzing the interaction of resonance with remote bodies positioned far away from resonators, where QNM fields largely diverge, as demonstrated in a recent study [89]. This study confirms that although QNM divergence leads to spectral instability, it does not cause any inconsistencies. The optical response of resonators disturbed by a distant body remains largely unchanged and can be accurately predicted using first-order QNM perturbation theory. It is also noteworthy that similar conclusions have been drawn in the context of gravitational waves [90-92].

## 3 Applications of QNMs in the Design of Optical Resonators

The essence of resonance design involves perturbing and hybridizing resonances to achieve specific optical responses. Traditionally, this process involves performing repeated full-wave simulations at real frequencies [2,28,49-52]. However, as mentioned in the introduction, this conventional approach encounters significant physical and computational limitations. In this section, we explore several examples to demonstrate how QNM methods overcome these challenges and facilitate the design of modern optical resonators, focusing first on QNM hybridization and then on QNM perturbation.

### 3.1 QNM hybridization

The motivations for mode hybridization are twofold.

First, hybridization allows us to combine the properties of different resonant modes to create new modes with distinct multifunctional properties [56-61,93-95], e.g. modes with both magnetic and electric responses [61,93,95-97], modes with both high radiative efficiency and small mode volume [2,56,58-60,98]. A well-known example is the hybridization of photonic microcavities and plasmonic resonator modes [99-101]. This approach utilizes plasmonic components to achieve deep subwavelength confinement (small mode volumes), while preserving a relatively high $Q$ factor, which is inherited from the photonic cavity.

Second, hybridization induces a chemical-like 'reaction' between modes, creating new modes with unique properties that go beyond a simple combination of the original modes. This serves as the second motivation for mode hybridization. For instance, when the eigenfrequencies of two QNMs cross each other as a parameter varies, the $Q$ factor of one mode can be significantly boosted reaching exceptionally high values [33,102,103]. Additionally, some studies have engineered interactions between modes to achieve exceptional points [104-107], where both the eigenvalues and eigenvectors of the interacting modes coalesce.

### 3.1.1 Mixing mode properties

A key feature of the QNM framework is its ability to compute the intrinsic properties of resonators and understand how these properties are affected by mode hybridization [56-60]. This capability allows for the design and optimization of nanoresonators, offering both computational efficiency and clear physical insights. This aspect is illustrated in Figure 2, with a structure known as a picocavity [2,108-110]. The latter consists of a NPoM, but additionally encompasses an individual atomic-scale protuberance on one of the gap surfaces, as shown in Figure 2a. This structure, which has recently garnered significant attention [2,56,57,108-110], demonstrates an impressive capability to confine light.

Picocavities exhibit two types of QNMs [56,57]. One type features an electric field that is highly localized near the protuberance, while the other exhibits electric field distributions akin to QNMs found in NPoM structures. We classify these as protuberance-like and NPoM-like QNMs, respectively. As depicted in Figure 2c, the protuberance-like QNM (mode II) provides extremely confined field characteristics, achieving a minimum mode volume below one cubic nanometer. However, due to the significant size disparity between the protuberance and the photon wavelength, this mode exhibits very low radiative efficiency. It cannot be efficiently excited by far-field sources (such as plane waves); when excited by a near-field emitter, almost all energy is converted into Ohmic losses in the metal, with minimal photon radiation in the far-field.

The mode volume of the NPoM-like mode (mode I) is much larger than that of the protuberance-like QNM. However, it offers high radiative efficiency and can be excited from the far-field. Its radiative efficiency, characterized by the intrinsic radiative diagram in Figure 2c, is approximately 20 times greater than that of the protuberance-like QNM.

A natural approach to achieve both high radiative efficiency and small mode volume involves coupling these two types of modes. The QNM framework facilitates monitoring the intrinsic properties as the modes hybridize. Figure 2b illustrates the eigenfrequencies of the hybridized QNMs as a function of the protuberance's aspect ratio, depicted by the red and blue solid curves. As the aspect ratio increases, the protuberance-like mode undergoes a redshift, and an anticrossing of resonant frequencies occurs when its eigenfrequency approaches the energy of the NPoM mode. In this region, the QNMs are significantly hybridized. As depicted in Figure 2c, both modes achieve a mode volume of less than 1 nm³ and demonstrate high radiative efficiency, combining the advantageous characteristics of both NPoM and protuberance modes.

### 3.1.2 Engineering the mode interaction

Another significant application of QNM theory lies in engineering mode interactions [102,111-113] to create new modes with properties distinct from a simple mixture of the original modes.

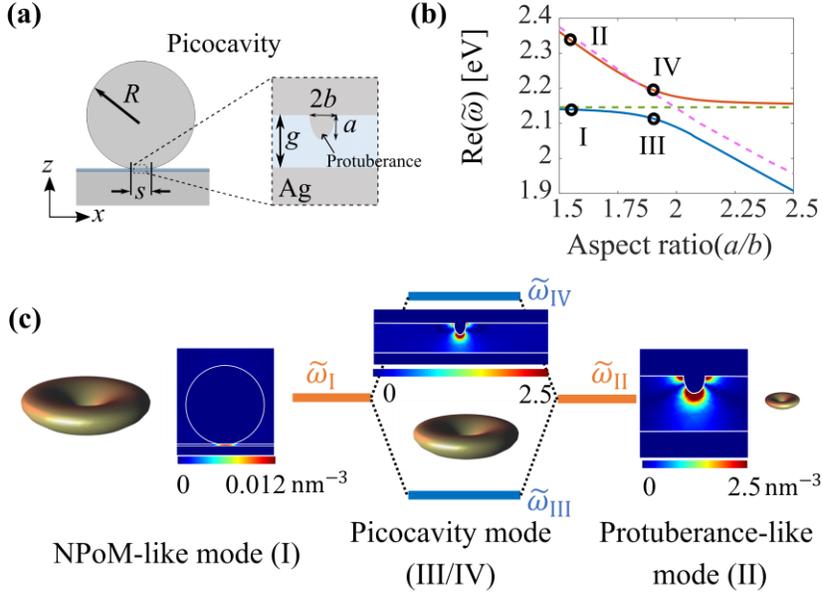

**Figure 2.** Mixing the properties of two QNMs by hybridization. (a) Schematic representation of a picocavity comprising an atomic-scale protuberance on a flat metal surface and a NPoM structure, each supporting a QNM. (b) Resonance frequencies as a function of the aspect ratio a/b of the protuberance. The mode of the protuberance strongly hybridizes with the mode of the NPoM for $a/b \approx 2$. (c) Energy level diagram illustrating the hybridization when the frequencies of the NPoM and protuberance modes are similar for $a/b = 2$. Outside the hybridization region, the structure exhibits an NPoM-like mode (QNM I in (b)) and a protuberance-like mode (QNM II in (b)). The NPoM-like mode has a relatively large mode volume and high radiative efficiency, while the protuberance-like mode features an ultra-small mode volume and a low radiative efficiency. The near-field maps depict the real part of the inverse mode volume $1/\tilde{V}_m$, while the far-field radiation diagrams illustrate how the normalized QNMs radiate in the far field. (a) and (b) are adapted from [56].

To illustrate our purpose, let us consider a resonator and let us deform it (Figure 3a). If we assume that two dominant QNMs are driving the resonator response, the new eigenfrequencies, $\tilde{\omega}_{hyb}$, of the deformed modes can be determined by solving a 2x2 eigenvalue problem [72]

$$\begin{pmatrix} \tilde{\omega}_1 & 0 \\ 0 & \tilde{\omega}_2 \end{pmatrix} \begin{pmatrix} a_1 \\ a_2 \end{pmatrix} = \tilde{\omega}_{hyb} \begin{pmatrix} 1 + V_{11} & V_{12} \\ V_{21} & 1 + V_{22} \end{pmatrix} \begin{pmatrix} a_1 \\ a_2 \end{pmatrix}, \tag{5}$$

where $\tilde{\omega}_{1(2)}$ are the QNM eigenfrequencies of the undeformed structure, $\tilde{\omega}_{hyb}$ is the unknown eigenfrequency of the deformed structure, and $a_1$ and $a_2$ are the modal excitation coefficients of the unperturbed QNMs. The terms $V_{ij}$ (with $i,j$=1, 2) are evaluated through an integral of the modal fields over the surface $S_r$ of the initial (undeformed) structure: $V_{ij} = \iint_{S_r} h \tilde{\mathbf{E}}_i^+ \cdot \Delta \boldsymbol{\varepsilon}(\tilde{\omega}_j) \tilde{\mathbf{E}}_j^- \, d^2\mathbf{r}$, where $h$ is the deformation perpendicular to the surface boundary (it varies with the curvilinear coordinate), and $\tilde{\mathbf{E}}_i^+$ and $\tilde{\mathbf{E}}_j^-$ the electric fields of the initial (undeformed) normalized QNMs at the outer (+) or inner (−) surface boundary. Outward deformations ($h > 0$) and inward deformation ($h < 0$) deformations (see Figure 3a) correspond to permittivity changes of $\Delta \boldsymbol{\varepsilon}$ and $-\Delta \boldsymbol{\varepsilon}$, respectively, where $\Delta \boldsymbol{\varepsilon} = \boldsymbol{\varepsilon}_{res} - \boldsymbol{\varepsilon}_b$ is the difference in permittivity between the resonator and background.

The $V_{ij}$ are generally complex numbers, with their imaginary parts arising from the non-Hermitian characteristics of the system. Typically, $V_{12} \neq V_{21}$, however if $\Delta\boldsymbol{\varepsilon}(\widetilde{\omega}_2) \approx \Delta\boldsymbol{\varepsilon}(\widetilde{\omega}_1)$, they are approximately equal.

The formula is derived using first-order approximations, ignoring contributions proportional to $h^2$ or higher-order terms for $V_{ij}$. Higher-order contributions can be included with more complex formulas. Nevertheless, except for specific cases where the first-order contribution vanishes due to symmetry, for instance, Equation (5) is accurate [72].

Using Equation (5), the new QNMs of the deformed structure are directly computed, enabling effective exploration of the parameter space to optimize the resonator shape. Figure 3b demonstrates the application of Equation (5) in designing nanoresonators that support exceptional points. This method significantly reduces the computation time needed to find the optimal design and provides insight into how geometric deformation influences mode hybridization.

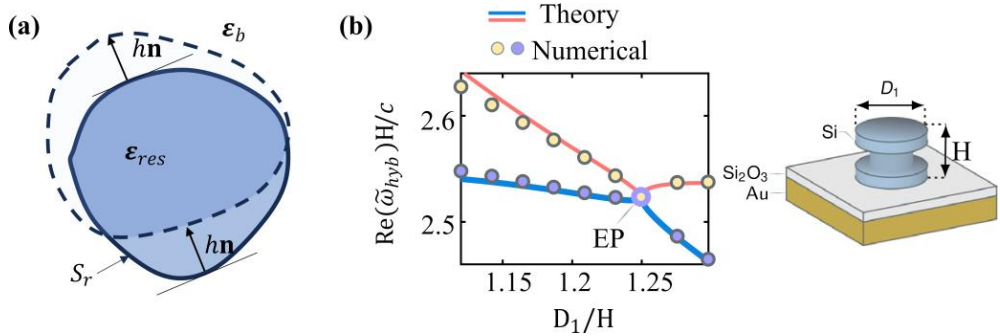

**Figure 3.** QNM interaction by deformation. (a) A nanoresonator is deformed with both inward and outward boundary changes. (b) Eigenfrequencies can be predicted with the analytical formula Equation (5) as we vary the shape. In (b), an exceptional point (denoted 'EP') is designed by deforming a nanoresonator. Full numerical calculations (circles) are compared to predictions obtained with Equation (5) (solid curves). Adapted from [72].

### 3.2 Mode perturbation

QNM perturbation theory of electromagnetic resonators is particularly useful for cavity design and has been used in many applications, including evaluating optical resonator sensitivity for optical biosensing, inverse design of high-Q optical cavities [114], and understanding or engineering the interplay of classical electromagnetism with other physical phenomena or nonlinear processes, such as thermo-optics [115], Kerr [116,117], or electron spill-out effects [43].

#### 3.2.1 Perturbation theory for understanding refractive index changes

Traditionally, designing optical sensors involves calculating numerous resonance spectrum variations for various perturbation instances (such as position, shape, and material), which is time-consuming when using a parametric frequency scan approach. Furthermore, for very small perturbations, simulations must achieve extremely high accuracy to ensure that signal changes caused by the perturbation are not obscured by numerical noise.

Small perturbations cause a small change of the complex frequency of all the QNMs, as illustrated in Figure 4a. The small variation is conveniently predicted using cavity perturbation theory.

The correct QNM normalization allows for deriving an accurate first-order perturbation formula for non-Hermitian systems [118]

$$\Delta\widetilde{\omega}_m = -\widetilde{\omega}_m \iiint_{\Omega_{per}} \Delta\varepsilon \tilde{\mathbf{E}}_m \cdot \tilde{\mathbf{E}}_m d^3\mathbf{r} + O(|\Delta\varepsilon|^2), \qquad (6)$$

where $\Omega_{per}$ is the finite volume of the perturber, and $\Delta\varepsilon = \varepsilon_{per} - \varepsilon_u$ is the difference in permittivity between the perturbed and unperturbed systems.

The predictive force of Equation (6) has been successfully validated for high-Q cavities by comparison with experimental data [119] and for low-Q plasmonic nanoresonators by comparison with full-wave computational data for various perturber shapes [118].

We may have noticed a difficulty also encountered in gravitational wave theory [80-84]: as the separation distance between the perturber and the resonator increases, the QNM field experienced by the perturber diverges. According to Equation (6), $\Delta\widetilde{\omega}_m$ should also diverge, contradicting our intuitive expectations that remote perturbers should not affect resonator characteristics. This apparent contradiction has been recently studied in detail [89]. The conclusion is clear: Equation (6) remains valid regardless of how far away the perturbations are.

The issue of remote perturbations and divergent coupling requires much care as it leads to spectral instabilities. The complex frequency plane becomes increasingly populated with numerous Fabry-Perot QNMs, and the optical response of the perturbed system is dominated by these Fabry-Perot QNMs instead of the initially perturbed one. Similar QNM instability issues are encountered in the gravitational wave physics of black holes [80-84,90-92]. The instability, also called 'the flea in the elephant effect', is caused by a small and remotely localized perturbation added to the black hole environment.

Equation (6) allows for the analytical evaluation of a crucial figure-of-merit (FoM) for optical resonators used in sensing applications, given by $\text{FoM} = S/\Delta\lambda$. Here $S$ represents the sensitivity, defined as the wavelength shift per unit change of the embedding medium refractive index, and $\Delta\lambda$ is the bandwidth of the resonance. Since $1/\Delta\lambda$ is proportional to the Q factor and $S$ is proportional to $\Delta\widetilde{\omega}_m$, from Equation (6), one can readily realize that $\text{FoM} \propto Q_m/\tilde{V}_m$, implying that an excellent candidate for an optical sensor should either possess a large $Q$ or a small mode volume.

The QNM perturbation toolbox for nanophotonic biosensor design has significantly expanded in recent years [41,120-123]. Equation (6) is valid for finite resonators perturbed by perturbers with finite size. This result has been extended to perturbations that cover the entire open space surrounding finite size resonators or periodic structures [121-123]. Equation (6) has also been extended to study systems that are perturbed by magnetic objects [120] or chiral molecules [41]. Refer to Figure 4 for more pictural details.

More recently, QNM perturbation theory has been further extended to predict the impact of perturbations on the optical scattering matrix [9,41]. This extension allows for the prediction of variations in optical responses, such as changes in spectrum intensity and lineshape, beyond just frequency shifts and linewidth changes. The theory has proven crucial in the design of nanophotonic sensors for chiral molecules [124,125]. It facilitates the rapid computation of the difference in circular dichroism spectra (ΔCD) with and without a chiral molecule. This is particularly important because, according to one of the authors [126], if the Pasteur parameter κ of the chiral molecule is extremely small, obtaining ΔCD can be computationally expensive due to the need for fine meshing to ensure the signal is above numerical noise.

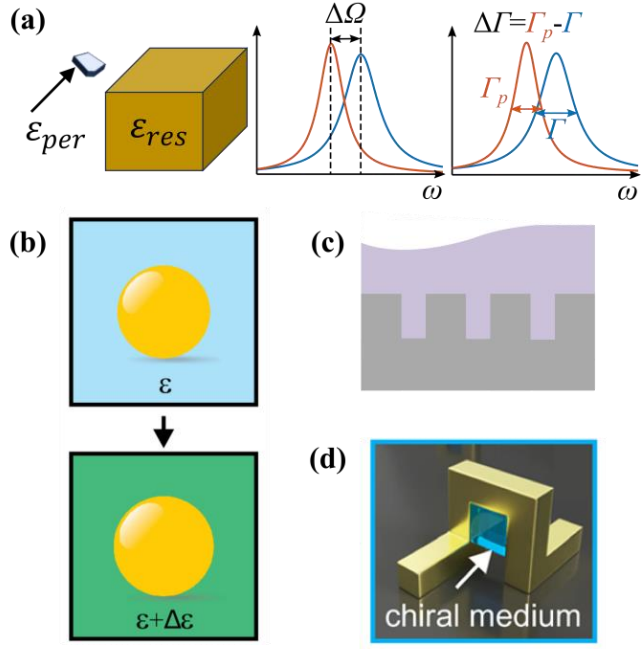

**Figure 4.** QNM perturbation theory for optical biosensors. (a) The introduction of a perturbation causes a resonant frequency shift ΔΩ and a linewidth change ΔΓ in the spectral response (Equation (6)) [118]. (b) The perturbation formula of Equation (6) has been extended for refractive-index changes extending over the entire open space surrounding the resonator [122]. (c) A perturbation formalism also exists for periodic structures [121]. (d) Perturbation formula has been developed for studying the index change caused by chiral mediums [41]. (b) is adapted from [122]; (c) is adapted from [121]; (d) is adapted from [41].

### 3.2.2 Inverse design of optical resonators

QNM perturbation theory can be especially useful for inverse design. Equation (6) is not accurate for this critical case, as boundary variations cause abrupt field changes inside the perturbation volume $\Omega_{per}$. This issue can be resolved using a technique known as local-field correction [127], which accounts for abrupt field changes by considering boundary conditions or the continuity of the electric field. By using local-field correction and assuming the perturbation does not cause hybridization between different QNMs, the frequency shift caused by shape deformation is given by $\Delta\widetilde{\omega}_m = -\widetilde{\omega}_m \iint_{S_r} h\widetilde{\mathbf{E}}_m^+ \cdot \Delta\boldsymbol{\varepsilon}(\widetilde{\omega}_m)\widetilde{\mathbf{E}}_m^- d^2\mathbf{r}$ [72,118], where the variables are the same as those used in Equation (5). In fact, Equation (5) reduces to the present formula in the absence of mode hybridization.

One recent application of the formula can be found in [114], where QNM perturbation theory was used in combination with a gradient-based algorithm to maximize the Q-factor of cavities formed in dielectric slabs with disordered nanoholes.

### 3.2.3 Quantum effect

QNM perturbation theory has also been successfully applied to understand the role of quantum effects in the response of nanoresonators with ultrasmall volumes, such as NPoM structures or picocavities. In these systems, strong confinement leads to non-classical effects, such as nonlocality and electron spill-out, which cannot be predicted by Maxwell's equations alone [128]. To accurately model these effects, numerical sampling must be significantly smaller than the Fermi wavelength, which is typically well below 1 nm [129]. This poses a significant challenge for incorporating quantum effects into classical Maxwell solvers.

QNM perturbation theory can potentially address this issue by treating non-classical effects as first-order perturbations of classical QNM fields. This approach allows for the analytical prediction of eigenfrequency changes due to non-classical corrections [43]. Recent advancements have further streamlined the computation of optical responses, such as Purcell factors and field enhancement factors [45]. They offer a quick method to evaluate how non-classical effects impact the optical properties of nanoresonators, including field confinement capability and scattering efficiency.

## 4      Perspectives and conclusion

Over the past decade, substantial advancements have been made in electromagnetic QNM theory, effectively addressing numerous critical challenges. The normalization of QNMs has been resolved [10-14], the completeness of QNM expansions has been confirmed for a variety of systems [10,20,21,74], and the physical implications and causes of QNM divergence are beginning to be understood [89].

These developments have facilitated the creation of various analytical QNM frameworks that significantly improve the design and comprehension of micro and nanoresonators. In this review, we have highlighted recent progress, showcasing their benefits in offering greater numerical efficiency and physical insights compared to traditional design approaches. We hope this will encourage a wider adoption of QNMs and further innovation in electromagnetism and other areas of physics.

Despite these successes, research on QNMs in electromagnetism continues vigorously and several open questions remain.

One of the foremost issues is the convergence of QNM reconstruction, as described in Equations (2) and (3). Achieving robust convergence is complex and influenced by numerous factors, including the material properties of the resonators [68], the choice of the formula for $\alpha_m$ [17,18], and the configuration of perfectly matched layers (PML) used for regularization [19]. Currently, the community lacks a definitive guideline on optimizing these parameters to improve convergence.

Another unresolved issue is understanding the existence of various QNM decomposition formulas. As discussed in Section 2.2, in dispersive systems, the formula for $\alpha_m$ is sensitive to the choice of auxiliary fields and source terms. Although all formulas share a resonant pole term $1/(\widetilde{\omega}_m - \omega)$, they differ in a non-resonant term $f(\omega)$, which is a slow-varying function of $\omega$. Recent studies have shown that certain choice of $f(\omega)$ leads to $\beta_m$ in Equation (3), derived from $\alpha_m$, exhibiting an instantaneous response term [27,130]. It would be important to verify that $f(\omega)$ for all the formulas offer a consistent physical interpretation.

From an application standpoint, there are numerous domains where QNM theory has yet to be fully utilized. For instance, QNM theory could potentially be applied to the analysis of spectra in photoemission electron microscopy (PEEM) [24], electron energy-loss spectroscopy (EELS), or high-order nonlinear optics [131], offering new insights and computational methods in these fields.

Finally, the recent interest in time-varying nanoresonators [132,133], whose optical properties can be modulated on time scales comparable to the oscillation period of electromagnetic fields, has opened up new avenues for QNM research. Extending the QNM framework to model the optical response of these dynamic, non-Hermitian systems could help discover a wide range of novel effects and applications.

## Author Contributions

TW and PL wrote the paper.

## Funding

## Acknowledgments

PL acknowledges financial support from the WHEEL (ANR-22CE24-0012-03) Project.